# The η/$s$ ratio in finite nuclei.

N. Auerbach

School of Physics and Astronomy, Tel Aviv University

Tel Aviv, 69978, Israel

and

S. Shlomo

Cyclotron Institute, Texas A&M University

College Station, TX  77843, USA

Abstract: Experiments at the Relativistic Heavy Ion Collider (RHIC) suggest that the state of matter produced in the experiments has a low shear-viscosity to entropy-density ratio η/$s$. We ask here the question what is this ratio in usual finite nuclei at low temperature. We use the experimental and theoretical results for the widths of giant vibrational states in nuclei in order to calculate the above ratio. We find that the values of η/$s$ are not very different from the ones found in the RHIC experiments.

In certain supersymmetric gauge theories one finds [1] that the ratio of shear viscosity η to entropy density $s$ is equal to:

$$\frac{\eta}{s} = \frac{\hbar}{4\pi k_B} \approx 6.05 \times 10^{-13} K\ s \qquad (1)$$

where $k_B$ is the Boltzmann constant. It has been conjectured that this ratio is the lower limit for a large class of quantum field theories. The analysis of the ultrarelativistic heavy ion collisions data from RHIC seems to indicate that the state of matter produced behaves like a liquid with the above ratio being close to the lower limit [2]. Thus the matter produced behaves as a perfect fluid.

It is however important to examine the values of the $\eta/s$ in systems close to the ones obtained in the relativistic collisions at RHIC. One such system is a finite nucleus. It is quite clear that there is strong affinity of the matter formed in the RHIC experiments and "conventional" finite nuclei. The same forces are active in both systems and after all the state in RHIC experiments is due to the interaction of finite nuclei. In this work we first demonstrate that a consistent value for $\eta$ is deduced from (i) analysis of the width of giant resonances within the hydrodynamic model, (ii) kinetic theory, and from (iii) the process of fission described using liquid drop models. We then provide a simple assessment of the entropy density.

The use of hydrodynamical models in nuclei has a long history. The first successful calculations of nuclear masses were obtained from a model that has considered nuclei to behave as liquid drops. Hydrodynamical models were employed in the description of vibrational states in nuclei. In particular so called giant resonances were described as vibrations of proton and neutron fluids [3]. The isoscalar vibrations consist of proton and neutron fluids collectively vibrating in phase, while the isovector ones are described as vibrations of the proton liquid out of phase, with the neutron fluid. In most cases these models were successful in reproducing the experimental excitation energies and cross sections of reactions exciting such resonances.

Most of the giant resonances, at excitation energies in the range of 10-40 MeV have a finite life time and carry a width. Following the success of the hydrodynamical models an attempt was made to link the widths of these resonances to the viscosity of the proton-neutron fluids [4]. In Ref. [4] a set of coupled hydrodynamical equations of the Navier- Stokes type was used to describe the flow of two viscous fluids, of protons and neutrons. The solution of these equations gave rise to giant resonances of both isoscalar (*I*=0) and isovector (*I*=1) type. For example the linearized Navier-Stokes equation for the isoscalar mode is:

$$\frac{\partial \vec{v}_0}{\partial t} = -\frac{1}{\rho} u_0^2 \vec{\nabla} \rho + \nu \nabla^2 \vec{v}_0 + \frac{1}{3} \nu \vec{\nabla} \vec{\nabla} \cdot \vec{v}_0 \qquad (2)$$

where: $\rho = \rho_n + \rho_p$ ($\rho_n$ and $\rho_p$ being the neutron and proton densities, respectively), $u_0$ is the wave velocity, $\nu$ the kinematical viscosity and

$$\vec{v}_0 = \frac{\rho_n \vec{v}_n + \rho_p \vec{v}_p}{\rho_n + \rho_p}. \qquad (3)$$

Solving this equation with appropriate boundary conditions and for various multipolarities one obtains eigenvalues containing a real and imaginary part. The real part represents the energy of the excitation and the imaginary part depends on the kinematical viscosity parameter $\nu$ and represents the lifetime of the excitation. The mass dependence A of the computed widths exhibits the experimental trends of the giant isoscalar resonances. As a result of such calculation one obtains a value for the kinematical viscosity [4]:

$$\nu = 0.6 \times 10^{22} \, \text{fm}^2 \, \text{sec}^{-1} \qquad (4)$$

The relation between the kinematical viscosity and the shear viscosity is:

$$\eta = \rho \nu \qquad (5)$$

For the above value of $\nu$ the shear viscosity is:

$$\eta \approx 1 \times 10^{-23} \, \text{MeV} \, \text{fm}^{-3} \, \text{sec} \qquad (6)$$

A later study also used a hydrodynamical model to study isoscalar giant resonances in nuclei and used the notion of viscosity to describe widths of these resonances [5].

The value of the shear viscosity coefficient deduced in this study was:

$$\eta \approx (1.9 \pm 0.6) \times 10^{-23} \, \text{MeV fm}^{-3} \, \text{sec} \qquad (7)$$

Recently, in Ref. [6], the authors described the dynamics of cold and hot nuclei within a generalized Fermi liquid drop model by employing a collision kinetic equation, which properly accounts for the dissipative propagation of sound waves in finite nuclei and nuclear matter. For a temperature $T < \varepsilon_F$ and excitation energy $\hbar\omega < \varepsilon_F$ of the sound wave, one finds for the collision viscosity

$$\eta = \frac{2}{5} \rho \varepsilon_F \frac{\tau_{coll}}{1 + (\omega \tau_{coll})^2} \qquad (8)$$

where

$$\tau_{coll} = \frac{\tau_0}{1 + (\hbar\omega / 2\pi T)^2} \qquad (9)$$

with

$$\tau_0 = \hbar\alpha / T^2 \qquad (10)$$

In Eqs. (8) - (9), $\tau_{coll}$ is the Landau approximation for the collision relaxation time deducted from the collision integral. The ω-dependent terms are due to memory effects, associated with the dynamical distortion of the particle momentum distribution. It was shown (see, for example Ref. [6]) that the form of Eq. (9) nicely describes both regimes of high and low frequencies ω, which correspond to zero sound (giant resonance) and first sound, respectively, as well as the intermediate regime. We note that in the low temperature limit of the collision relaxation time $\tau_{coll}$ it is very important to take into account its dependence on ω, for high ω. It is seen from Eq. (9) that in the rare collisions zero sound regime (ωτ >> 1, T << ℏω) we have that $\tau_{coll} \sim 1 / (\hbar\omega)^2$ (with a weak T dependence) and in the frequent collisions first sound regime (ωτ << 1, T >> ℏω), we have $\tau_{coll} \sim 1 / T^2$. The corresponding value of η is then given by $\eta = \frac{2}{5}\rho\varepsilon_F \frac{\hbar}{4\pi^2\alpha}\left[1+(2\pi T/\hbar\omega)^2\right]$ and $\eta = \frac{2}{5}\rho\varepsilon_F \hbar\alpha/T^2$, for the zero and first sound regimes, respectively (see also Eqs. (388) and (385) of Ref. [6], respectively, and Eqs. (10.22) and (10.12) of Ref. [7], respectively). The value of α in Eq. (10) is sensitive to the in-medium-nucleon-nucleon scattering cross section. Taking the in-medium cross-section to be ½ of the free nucleon-nucleon cross section, one finds [6] that $\alpha = 9.2$ MeV. For illustration we show in Fig. 1a the value of η as a function of the temperature T, obtained using the values of $\varepsilon_F = 40$ MeV, $\rho = 0.16$ fm$^{-3}$, $\alpha = 9.2$ MeV and $\hbar\omega = 20$ MeV. Note that our results are relevant for temperature $0.5$ MeV $<$ T $< 5$ MeV, were giant resonances exist.

At low temperature, $T \ll \varepsilon_F$, the relation between the thermal excitation energy $E^*$ and T is given by

$$E^* = aT^2 \qquad (11)$$

where $a$ is the level density parameter obtained from [8]

$$a = \frac{(1+\beta)A\pi^2}{6\varepsilon_F} \qquad (12)$$

with A being the number of nucleons. For an ideal Fermi gas we have that $\beta = 1/2$. However, for non interacting nucleons in a finite potential well (Woods-Saxon) we have that $\beta = 1$ and $\beta = 2$ for a heavy nucleus (A~200) and a light nucleus (A~20), respectively [8]. For an ideal Fermi gas, for $A = 200$ and a thermal excitation energy $E^* \approx \hbar\omega = 20$ MeV, one obtains $T \approx 1$ MeV which then, using Eqs. (8 - 12), gives:

$$\eta \approx 0.5 \times 10^{-23} \text{ MeV fm}^{-3} \text{ sec} \qquad (13)$$

The three values for η quoted here were obtained by considering the dissipation of collective motion as exhibited by the giant resonances. The values deduced in all three cases are consistent and the range of the values is:

$$\eta \approx (0.5 - 2.5) \times 10^{-23} \, \text{MeV fm}^{-3} \, \text{sec} \qquad (14)$$

Another type of collective motion encountered in the dynamics of nuclei is the process of fission. A number of works appeared in the literature which dealt with the dynamics of fission in heavy nuclei using viscous liquid drop models [9, 10, 11]. For example, in [9] the authors use a macroscopic approach and solve classical equations of motion for the fissioning nucleus. They apply this to spontaneous and induced fission. They find that the average value for the shear viscosity that reproduces best the data is:

$$\eta \approx (0.9 \pm 0.3) \times 10^{-23} \, \text{MeV fm}^{-3} \, \text{sec} \qquad (15)$$

In a later study [10], applying a somewhat different model, the value deduced is twice as large:

$$\eta \approx (1.9 \pm 0.6) \times 10^{-23} \, \text{MeV fm}^{-3} \, \text{sec} \qquad (16)$$

In another work [11] that studied nuclear fission the value obtained for $\eta$ was:

$$\eta \approx 1 \times 10^{-23} \, \text{MeV fm}^{-3} \, \text{sec} \qquad (17)$$

The values of η found in the studies of the fission process agree generally with the ones found from the work on giant resonances. We will therefore use for the shear viscosity the range of values obtained from the exploration of the widths of giant resonances in Refs. [4] and [5]. The results from the other approaches are in good agreement with these values.

In order to evaluate the ratio in Eq. (1) one needs to evaluate the entropy. Our first, simplest determination of the nuclear entropy will be based on models of a free

Fermi gas or of non interacting nucleons contained in an average nuclear potential, such as a finite Woods-Saxon well. For low temperature, $T < \varepsilon_F$, the entropy $S$ of such a system is given by the simple expression [6, 8, 12]:

$$S = 2aT \tag{18}$$

where $a$ is the level density parameter, Eq. (12). Here $T$ and $\varepsilon_F$ are in MeV and S is in units of the Boltzmann constant $k_B$. Transforming $S$ to the entropy density $s$ we have:

$$s = \frac{\rho}{A} S \tag{19}$$

In figure 1b we show $s$ as a function of $T$.

Using Eqs. (18), (19) and (12), we find, for $\beta = 1/2$, that:

$$\frac{\eta}{s} \approx (5-25) \times \frac{\hbar}{4\pi k_B} \tag{20a}$$

This is for a free Fermi gas. For a gas of nucleons moving in a Woods-Saxon potential we have for a large and a small nucleus the corresponding values:

$$\frac{\eta}{s} \approx (4-19) \times \frac{\hbar}{4\pi k_B} \quad \text{and} \quad \frac{\eta}{s} \approx (2.5-12.5) \times \frac{\hbar}{4\pi k_B} \tag{20b}$$

Combining the equations (8)-(10) for temperature dependence of the shear viscosity with the equation of the entropy density, Eq. (19), we can write for an ideal Fermi gas (see details in Refs. [6, 8]):

$$\frac{\eta}{s} = \frac{\hbar}{4\pi k_B} \frac{16}{5\pi} \varepsilon_F^2 \frac{\alpha}{T} \frac{T^2 + (\hbar\omega/2\pi)^2}{(\hbar\omega\alpha)^2 + \left[T^2 + (\hbar\omega/2\pi)^2\right]^2} \tag{21}$$

In Fig.1c we show $\eta/s$ as a function of temperature. The curve has the characteristic behavior found in other fluids [1], the minimum occurring at $T = 3$ MeV, with the value of $\eta/s \approx 3 \times \hbar/4\pi k_B$.

We add that in the rare collision regime ($\omega\tau \gg 1$), at low temperature $T \ll \hbar\omega$, one finds from Eq. (21) that

$$\frac{\eta}{s} = \frac{\hbar}{4\pi k_B} \frac{4}{5\pi^3} \varepsilon_F^2 \frac{1}{\alpha T} \tag{22}$$

The 1/T behavior of $\eta/s$ is clearly seen in Figure 1c for T≤1 MeV. In the frequent collision regime ($\omega\tau \ll 1$), at high temperature $T \gg \hbar\omega$ (not shown in the figure), we have:

$$\frac{\eta}{s} = \frac{\hbar}{4\pi k_B} \frac{16}{5\pi} \left(\frac{\varepsilon_F}{T}\right)^2 \frac{\alpha}{T} \tag{23}$$

It is intriguing to note that in medium energy heavy ion collisions one may create a hot and dense nuclear matter with a density of several times the saturation density $\rho_0$, i.e. nuclear matter with $\varepsilon_F \approx$ 100 MeV. For $T \approx \varepsilon_F / 2$ one obtains from Eq. (23) values for $\eta/s$ which are close to $\hbar/4\pi k_B$.

The experiments from RHIC provide data that can be used to determine the $\eta/s$ ratio of the state created in these super-relativistic collisions of heavy nuclei. It is presently widely accepted [2] that the form of matter created has a very high fluidity with:

$$\frac{\eta}{s} \leq 5 \times \frac{\hbar}{4\pi k_B} \tag{24}$$

Comparing this value to our results for finite nuclei at low temperature, Eqs. (20a, 20b), we see that the deduced ratio (especially the lower limit) is not drastically different from the RHIC result. It is possible that the strong fluidity is a characteristic feature of the strong interaction of the many-body nuclear systems in general and not just of the state created in the relativistic collisions. More studies are required in order to understand this point.

**Acknowledgements**

N. Auerbach thanks B. Jennings and A. Schwenk for their hospitality at TRIUMF where part of this work was performed. This research was supported in part by the US Department of Energy under grant No. FG03-93ER40773. S. Shlomo thanks the School of Physics and Astronomy of Tel Aviv University for the kind hospitality

**Figure Caption**

Figure 1. The nuclear shear viscosity $\eta$ (a), entropy density *s* (b) and their ratio, ($\eta/s$) (c), in units of $\hbar/4\pi k_B$, as functions of the temperature, T. The values of the parameters used in the calculations are; $\varepsilon_F = 40$ MeV, $\rho = 0.16$ fm$^{-3}$, $\alpha = 9.2$ MeV and $\hbar\omega = 20$ MeV.

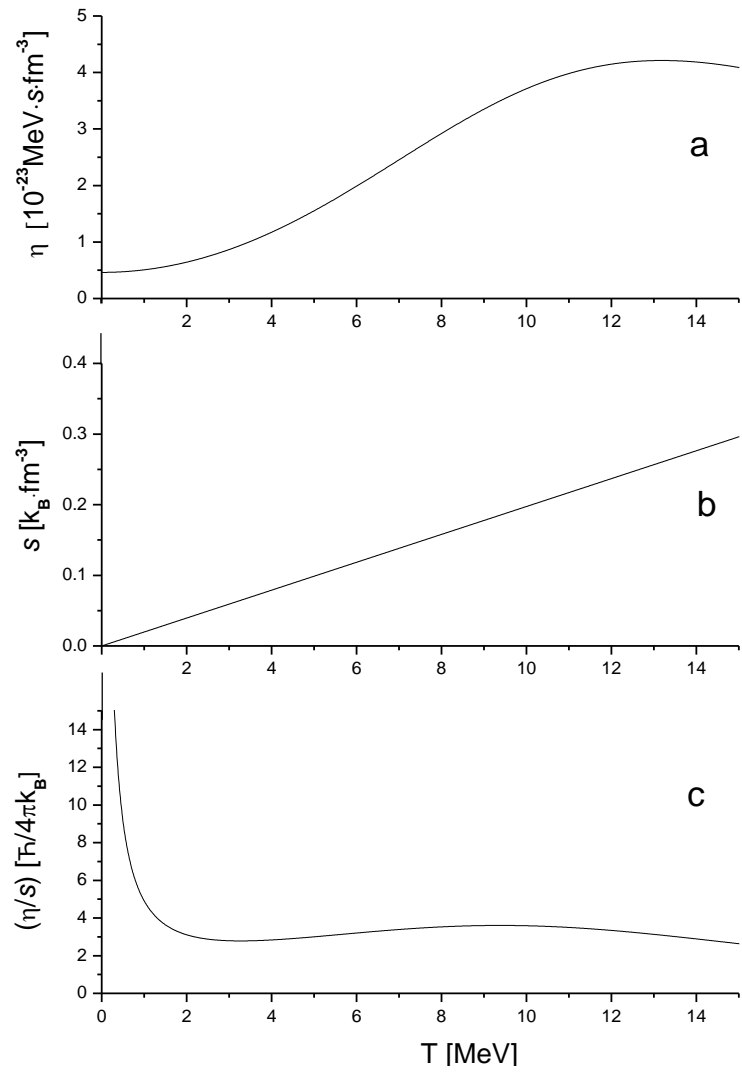

Figure 1